\documentclass[a4paper]{jpconf}
\usepackage{bm}
\usepackage{graphicx}

\newcommand{\be}{\begin{eqnarray}}
\newcommand{\ee}{\end{eqnarray}}
\newcommand{\no}{\nonumber}

\begin{document}
\title{Zeros of the partition function and dynamical singularities in spin-glass systems}

\author{K Takahashi$^1$ and T Obuchi$^{2,3}$}

\address{${}^1$ Department of Physics, 
Tokyo Institute of Technology, Tokyo 152-8551, Japan}
\address{${}^2$ Cybermedia Center, Osaka University, Osaka 560-0043, Japan}
\address{${}^3$ CNRS-Laboratoire de Physique The\'orique de l'ENS, 24 rue Lhomond, 75005 Paris, France}


\begin{abstract}
We study spin-glass systems characterized by 
continuous occurrence of singularities.
The theory of Lee-Yang zeros is used to find the singularities.
By using the replica method in mean-field systems, we show 
that two-dimensional distributions of zeros of the partition function 
in a complex parameter plane are  
characteristic feature of random systems.
The results of several models indicate that 
the concept of chaos in the spin-glass state 
is different from that of the replica symmetry breaking.
We discuss that a chaotic phase at imaginary temperature 
is different from the spin-glass phase and 
is accessible by quantum dynamics in a quenching protocol.
\end{abstract}

\section{Introduction}

Phase transition is one of the most interesting phenomena 
in statistical physics and has been studied in various ways.
The theory of spin glasses has attracted interests for many years 
as it is important not only for describing real materials
but also for realizing 
the unreasonable effectiveness of the statistical mechanics.

Previous works revealed that 
the peculiarity of the spin-glass state is attributed to 
a rugged structure of the free energy landscape~\cite{MPV, Nishimori}.
There exist many pure states in the configuration space 
separated by infinite barriers  
and the system takes one of their states.
By sweeping a parameter such as the temperature,
we can find continuous changes of the state,
which means that a certain kind of  phase transitions occur continuously.

Phase transition is generally expected to involve singularities 
in thermodynamic functions.
However, it is known for example in the analytical result of 
the Sherrington-Kirkpatrick model~\cite{SK} 
that the free energy is a continuous function with respect to 
the temperature in the spin-glass phase~\cite{Parisi1, Parisi2}.
Although the model exhibits a continuous change of the state 
known as the temperature chaos~\cite{RC}, 
such a property cannot be seen in the free energy.
This is not a surprising result since a single singularity is hidden 
in continuous singularities.
Thus, from a specific thermodynamic function, 
one cannot judge whether there exist continuous singularities. 

Lee and Yang proposed a theoretical method to find 
the singularity in the thermodynamic functions
in a rather primitive way~\cite{YL, LY}.
They discussed that the singularity is identified 
by the zeros of the partition function 
and showed explicitly that 
the phase transition in the Ising model is described by the zeros.
This theory is suitable to our purpose to see the singularity directly.
The other advantage of this method is that 
zeros exist even in finite systems and 
one can study how the system approaches the thermodynamic limit.
We can also argue that the role of zeros is 
not only in finding the phase transition point
but also in calculating any thermodynamic functions.

In this paper, according to our previous works,
we develop the theory of zeros in spin-glass systems.
We use a reliable systematic method to calculate 
the zeros of the partition function and 
discuss the advantage of using zeros for studying spin-glass transitions 
and the related peculiar behaviours~\cite{Takahashi, OT1}.
In order to discuss further possibilities of exploring zeros,
we also consider some application 
on dynamical properties of the systems~\cite{OT2}.

\section{Zeros of the partition function}
\label{sec:zero}

\subsection{Theory of zeros}

In a finite system, the partition function $Z=\Tr\exp(-\beta H)$,  
where $H$ is the Hamiltonian and $\beta$ the inverse temperature, 
is an analytic function with respect to a parameter $y$ 
such as $\beta$ and the magnetic field.
It can be written in a factorized form 
\be
 Z(y)= {\mathrm e}^{C}\prod_i(y-y(i))^{\eta_i},
\ee
where $y(i)$ is $i$th zeros, $\eta_{i}$ is the positive power index and 
the factor $C$ is an irrelevant constant.
Since the partition function is positive by definition, 
$y(i)$ must be outside the domain where the parameter $y$ is defined.
Otherwise, we have an unacceptable result that 
the partition function becomes zero at $y=y(i)$.
Therefore, $y(i)$ is a complex number in general.
The free energy density $f$ is extracted from the partition function
as $-\beta f=(1/N)\ln Z$ where $N$ is the system size.
By using the zeros of the partition function $\{y(i)\}$, we can write 
\be
 -\beta f(y)=\frac{C}{N}
 +\int {\mathrm d}z_1{\mathrm d}z_2\,\rho(z_1,z_2)\ln (y-z),
 \label{f}
\ee
where $z$ is complex as $z=z_1+iz_2$ and the density of zeros is defined as 
\be
 \rho(z_1,z_2)=\frac{1}{N}\sum_i\eta_i
 \delta(z_1-{\rm Re}\,y(i))\delta(z_2-{\rm Im}\,y(i)).
\ee
The power $\eta_{i}$ now plays a weight of the corresponding zero.
Thus, we can find the free energy by knowing the density of zeros.
All of properties on the thermodynamic state are reflected in zeros.
This can also be understood from the fact that 
the density of zeros is calculated from the partition function
with complex $y$ as 
\be
 \rho(y_1,y_2)=\frac{1}{2\pi N}\left(
 \frac{\partial^2}{\partial y_1^2}
 +\frac{\partial^2}{\partial y_2^2}\right)
 \ln |Z(y=y_1+iy_2)|. \label{lnz}
\ee

Since we can understand all thermodynamic properties by zeros, 
the phase transition is also found from their distribution.
As we mentioned above, in finite systems, the parameter $y$ never 
coincides with any of $y(i)$ as long as $y$ takes real physical values. 
However, this is not the case for systems at the thermodynamic limit.
If we find a singularity in a thermodynamic function, 
it must be described by the zeros of the partition function.
This implies that the zeros asymptotically approach the physical domain.
The location of the phase transition point must be identified by the zeros.

\subsection{Replica method}

We treat random systems and take an average
over random realizations of parameters in the Hamiltonian.
In order to calculate the density of zeros, 
we use the formula in equation (\ref{lnz}).
We must calculate the random average of the logarithm of 
the partition function.
It is known in spin-glass theory that the replica method 
is useful for such a calculation.
Since we must treat the absolute value of the complex partition function,
we use the modified formula 
\be
 \ln |Z|=\lim_{n\to 0}\frac{|Z|^{2n}-1}{2n}.
\ee
$2n$-replicated systems are prepared to calculate the density of zeros.
This is the main idea in our analysis.

In order to see how the density of zeros is calculated, 
we sketch the idea by using the Edwards-Anderson model 
\be
 H = -\sum_{\langle ij\rangle}J_{ij}S_{i}S_{j}, \label{EA}
\ee
of Ising spins $S_i$ on a lattice~\cite{EA}.
Here, we take the summation over pairs of spins 
and $J_{ij}$ is the Gaussian random variable with the variance  
$[J_{ij}^2]=J^2$.
We study zeros in the complex-$\beta J$ plane~\cite{Fisher}.
The average of the $2n$-replicated partition function 
is calculated as 
\be
 [|Z|^{2n}] &=& \left[\Tr\exp\left\{
 \beta\sum_{a=1}^n \sum_{\langle ij\rangle}J_{ij}S_i^aS_j^a
 +\beta^*\sum_{a=1}^n \sum_{\langle ij\rangle}J_{ij}{S^*_i}^a{S^*_j}^a
 \right\}\right] \no\\
 &=& \Tr\exp\left\{
 \frac{1}{2}\sum_{a,b}^n \sum_{\langle ij\rangle}
 \left(\beta^2J^2 Q_i^{ab}Q_j^{ab}
 +{\beta^*}^2J^2{Q^*_i}^{ab}{Q^*_j}^{ab}
 +2\beta\beta^*J^2 \tilde{Q}_i^{ab}\tilde{Q}_j^{ab}
 \right)
 \right\}, \label{Z}
\ee
where the trace is taken over Ising spin variables 
$\{S_i^a\}_{i=1,\cdots, N}^{a=1,\cdots, n}$ and 
$\{{S^*_i}^a\}_{i=1,\cdots, N}^{a=1,\cdots, n}$.
We also put 
\be
 Q_i^{ab}=S_i^aS_i^b, \qquad
 {Q^*_i}^{ab}={S^*_i}^a{S^*_i}^b, \qquad
 \tilde{Q}_i^{ab}=S_i^a{S^*_i}^b.
\ee
These variables are identified with the order parameters of the system.
By taking the thermodynamic limit $N\to\infty$, we write 
the large deviation form 
$[|Z|^{2n}] =\exp\left(N\phi_n(\beta,\beta^*)\right)$
and the logarithm of the partition function is calculated as 
\be
 \frac{1}{N}[\ln|Z|] = \lim_{n\to 0}\frac{1}{2n}\phi_n(\beta,\beta^*).
\ee
The calculation goes along the same way 
as the standard analysis.
Since it is difficult to perform the calculation in finite dimensional
systems, we mainly study infinite-range models where 
the mean-field method is applicable.
The additional difficulty in the present case lies in the fact that 
the function incorporates the complex parameters.

What can we expect for the result?
Generally, the result will be written in a form 
\be
 \lim_{n\to 0}\frac{1}{2n}\phi_n(\beta,\beta^*)
 =\phi^{(1)}(\beta)+\phi^{(1)}(\beta^*)
 +\tilde{\phi}(\beta,\beta^*).
\ee
$\tilde{\phi}$ cannot be written as a function of $\beta$ or $\beta^*$
and depends on $|\beta|^2=\beta\beta^*$.
In this case, only the last term contributes to 
the density of zeros:
\be
 \rho(\beta_1,\beta_2)=\frac{1}{2\pi}\left(
 \frac{\partial^2}{\partial\beta_1^2}
 +\frac{\partial^2}{\partial\beta_2^2}\right)
 \tilde{\phi}(\beta,\beta^*).
\ee
The density of zeros vanishes if this term is absent.
We see from equation (\ref{Z}) that 
there should exist the correlation $\tilde{Q}_{i}^{ab}$ 
to obtain a finite density of zeros.
By noting that such a correlation never occurs in pure systems,
we see that the two-dimensional distribution of zeros  
is specific to random systems.

We note the other possibility for obtaining finite density of zeros.
When the phase transition occurs,
we have a discontinuous behaviour of the thermodynamic functions.
Although $\phi_n(\beta,\beta^*)$ is expected to be continuous everywhere, 
the discontinuity can appear in its derivative.
In that case, the zeros exists on phase boundaries  
in the complex parameter space.
An example is known in the Ising model as the circle theorem~\cite{LY}.

\subsection{Possible patterns of pure state}

\begin{figure}[t]
\begin{center}
\includegraphics[width=18pc]{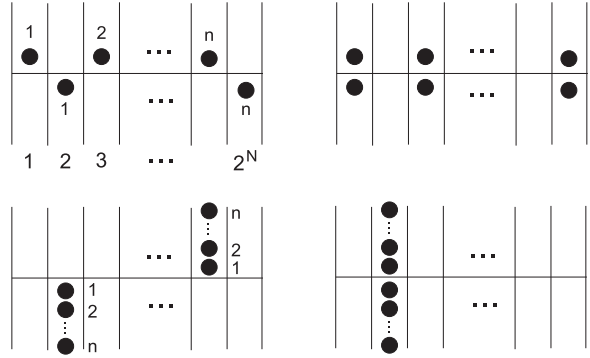}
\caption{\label{fig:rs}
Replica-symmetric configurations of pure states.
Each configuration has two kinds of boxes which represent
the normal space and the complex-conjugated one respectively. 
The upper diagrams represent paramagnetic phases and lower spin-glass phases.
$\tilde{Q}=0$ in the left diagrams and $\tilde{Q}\ne 0$ in the right.}
\end{center}
\end{figure}

\begin{figure}[t]
\begin{center}
\includegraphics[width=18pc]{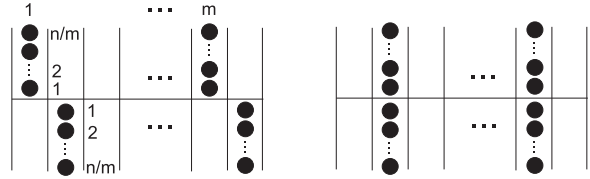}
\caption{\label{fig:1rsb}
Configurations in one-step replica-symmetry-breaking spin-glass phases.
$\tilde{Q}=0$ in the left diagram and $\tilde{Q}\ne 0$ in the right.}
\end{center}
\end{figure}

A thermodynamic state is described by a typical configuration 
of the microscopic variables.
This is reflected in the structure of the order parameter functions.
In random systems, we have correlations between replicas, and 
we can consider possible patterns of replica symmetry breaking generally.
This can be considered schematically in the following way.
We represent each replica by a ball and 
each configuration (pure state) by a box.
A typical state is represented by arranging balls in boxes.
We can consider various types of configurations, and typical 
states are summarized in figures~\ref{fig:rs} and \ref{fig:1rsb}. 
Since we consider complex-conjugated systems, 
we have two types of boxes, each has $2^N$-boxes.
They have parameters $\beta$ and $\beta^*$ respectively.
$n$-balls are arranged in a one type of boxes 
and the other $n$-balls are arranged in the other type of boxes.

In a replica symmetric state, the order parameter is represented by
$Q^{ab}=Q$ and ${Q^*}^{ab}=Q^*$ where $a\ne b$.
In the upper diagrams of figure~\ref{fig:rs}, 
a single ball is in a box at most and 
$Q=Q^*=0$, which means that 
the corresponding states are paramagnetic states. 
In the upper-left diagram of the figure,
$\tilde{Q}^{ab}=0$ which means that there is no correlation
between conjugated pairs.
We obtain $[|Z|^{2n}]=[Z^n][{Z^*}^n]$ and 
the density of zeros vanishes as a result.
On the other hand, for the upper-right type, 
$\tilde{Q}^{ab}\ne 0$ and we have a finite density of zeros.
Thus, we can define two kinds of paramagnetic state in the present analysis.
On the other hand, the replica-symmetric spin-glass state 
is specified by the lower two figures.
All balls are inside a single box and 
the correlations between different replicas
are finite, which imply a spin-glass state with $Q\ne 0$ and $Q^*\ne 0$.
The replica-symmetric spin-glass state appears 
in several simple models such as the spherical model.

In a similar way, we can consider spin-glass phases with $Q^{ab}\ne 0$
and ${Q^*}^{ab}\ne 0$ represented by 
the one-step replica-symmetry-breaking
solutions in figure~\ref{fig:1rsb}.
We have two possibilities also in this case 
with zero- and finite- $\tilde{Q}$.
The breaking parameter $m$ is determined by the equation of state 
and depends on the replica number $n$ and other parameters such as $\beta$.
These constructions of possible solutions are further generalized 
to cases of higher-step replica-symmetry-breaking states.
In this paper, we only treat systems described by
replica-symmetric and one-step replica-symmetry-breaking solutions
for simplicity.

The important point to notice here is that states with $\tilde{Q}\ne 0$
can appear in both the paramagnetic and spin-glass phases.
Since it is not clear whether 
such phases appear in models to be considered, 
we give explicit analysis in the following.

\subsection{Zeros and chaos}

We examine mean-field models exhibiting spin-glass phase transitions.
All of our systems are described by the $p$-spin Hamiltonian
\be 
 H = -\sum_{i_1<i_2<\cdots<i_p}J_{i_1i_2\cdots i_p}S_{i_1}S_{i_2}\cdots S_{i_p},
\ee
where the random averaging is taken by the Gaussian distribution
\be
 {\rm Prob}(J_{i_1i_2\cdots i_p})
 =\sqrt{\frac{N^{p-1}}{\pi J^2p!}}\exp\left(-\frac{N^{p-1}}{J^2p!}
 J_{i_1i_2\cdots i_p}^2\right).
\ee
Using the standard strategy for mean-field systems~\cite{MPV, Nishimori}, 
we can arrive at 
\be
 & & \phi_n(\beta,\beta^*) =
 -\frac{p-1}{4}\sum_{a,b}^n
 \left(\beta^2J^2(Q_{ab})^p+{\beta^*}^2J^2(Q_{ab}^*)^p
 +2|\beta|^2J^2(\tilde{Q}_{ab})^p\right)
 \no\\
 & & 
 +\ln\Tr\exp\left\{\frac{p}{4}\sum_{a,b}^n\left(
 \beta^2J^2(Q_{ab})^{p-1}S^aS^b
 +{\beta^*}^2J^2(Q^*_{ab})^{p-1}{S^*}^a{S^*}^b
 +2|\beta|^2J^2(\tilde{Q}_{ab})^{p-1}S^a{S^*}^b
 \right)\right\}, \no\\
 \label{phiq}
\ee
where the order parameters $Q$, $Q^*$ and $\tilde{Q}$
are determined by the condition to extremize $\phi_n(\beta,\beta^*)$.
We impose several ansatz on those parameters 
to solve the saddle-point equations.

\begin{figure}[t]
\begin{center}
\includegraphics[width=18pc]{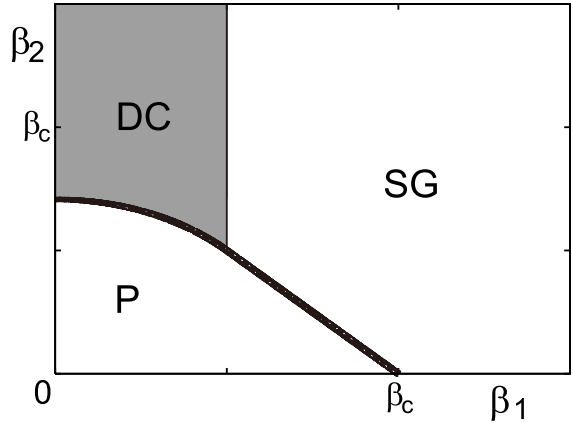}
\caption{\label{fig:rem}The distribution of zeros in the random energy model.
Zeros distribute in the dynamical chaotic (DC) phase uniformly
and on the phase boundary of the paramagnetic (P) phase
as shown by shaded area and bold line respectively.
There are no zeros in the spin-glass (SG) phase.
The critical point is given by $\beta_{\rm c}=2\sqrt{\ln 2}/J$.
}
\end{center}
\end{figure}

\begin{figure}[t]
\begin{center}
\includegraphics[width=16pc]{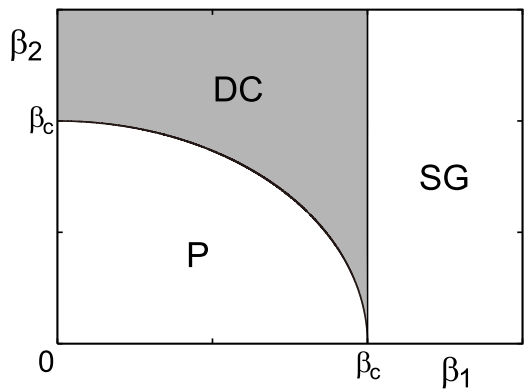}
\caption{\label{fig:sph1}
The distribution of zeros in the spherical model at $p=2$.
Zeros distribute uniformly in the DC phase.
The critical point is given by $\beta_{\rm c}=1/J$.
}
\end{center}
\end{figure}

In order to compare several results, 
we study two solvable models: the random energy model and spherical model.
The random energy model is defined by the Ising spin $S_i^a=\pm 1$ 
with the many-spin interaction limit $p\to\infty$~\cite{Derrida1, Derrida2}. 
The spin-glass phase is described by 
the one-step replica-symmetry-breaking solution.
In the spherical model, the spin variable is real and the constraint
$\sum_{i=1}^N(S_i^a)^2=N$ is imposed.
When $p=2$, the model has the replica-symmetric spin-glass phase
and for $p>2$ the one-step replica-symmetry-breaking one.
We show the result of the random energy model and 
spherical model with $p=2$ 
in figure~\ref{fig:rem} and \ref{fig:sph1} respectively.
The details are studied in~\cite{Takahashi, OT1}.

\begin{figure}[t]
\begin{center}
\includegraphics[width=16pc]{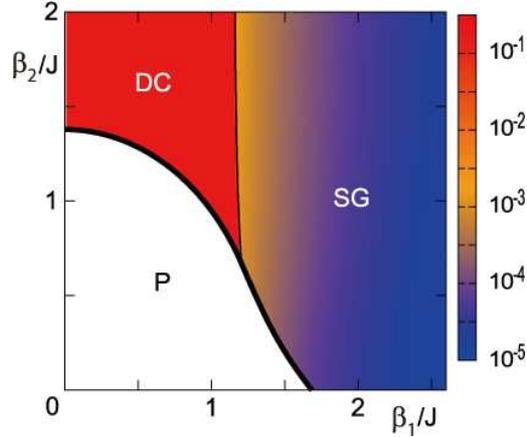}
\caption{\label{fig:sph2}The distribution of zeros in the 
multiple-interacting spherical model with $p=3$ and 4.
Zeros distribute in the SG and DC phases 
with values of the density denoted in colour and
on the phase boundary of the P phase.}
\end{center}
\end{figure}

In both the results, zeros 
appear in a paramagnetic phase (denoted by DC), characterized 
by the upper right diagram in figure~\ref{fig:rs}, and 
not in the spin-glass phase.
The spin-glass phases are described by the left diagrams
in figures~\ref{fig:rs} and \ref{fig:1rsb}.
Though $\tilde{Q}\ne 0$, we do not see any zeros in the spin-glass phases.
The same conclusion is found also in the spherical models with $p>2$.
We do not blame this result on the oversimplified patterns of 
the replica symmetry breaking.
In fact, for the random energy model with complex magnetic field, 
the zeros appear in the spin-glass phase~\cite{MP2, MP3}.
The difference lies in the chaotic nature of the spin-glass 
state~\cite{FPV, FN, Rizzo1, Rizzo2, RC, RY}.
In the spin-glass phase of the random energy model,
the system falls into the ground state and nothing spectacular
occurs even if we lower the temperature.
However, if we change the magnetic field, the pure state
changes accordingly.
In that case, 
the state is highly sensitive to the parameter variation.
We expect that such a chaotic nature is more important than
the step number of the replica symmetry breaking.

To confirm this speculation, we examine the spherical model 
with multiple interactions~\cite{RY}.
We consider the sum of the Hamiltonian with $p=3$ and $p=4$.
This model is known to show the temperature chaos effect.
We numerically solve the saddle-point equations to draw 
the phase diagram in figure~\ref{fig:sph2}.
The spin-glass phase is described 
by the one-step replica-symmetry-breaking,
which is the same as the standard spherical model at $p>2$.
However, in this case, the zeros appear in the spin-glass state
which means the relevance of the chaotic effect for the distribution
of zeros.

\section{Dynamical singularity}
\label{sec:dynamics}

\subsection{Quantum quench}

Our formulation shows that $\tilde{Q}\ne 0$ is necessary to find 
two-dimensional distributions of zeros.
We discussed in the previous section 
the chaotic (and nonchaotic) nature of the spin-glass phase
with $Q\ne 0$ and $Q^*\ne 0$.
On the other hand, we also find a different chaotic 
non-spin-glass phase with $\tilde{Q}\ne 0$ and $Q=Q^*=0$.
The analysis of several systems shows that such a phase
appears around the imaginary axis in the complex temperature plane.
It does not approach the real axis and cannot describe 
any thermodynamic phase transitions.

Our idea of exploring this new kind of chaotic phase
is to consider real time dynamics.
Since the Boltzmann factor with imaginary temperature corresponds
to the time evolution operator in quantum mechanics, 
we expect that the chaotic phase is relevant to quantum dynamics.

In order to introduce the quantum nature, we consider 
quantum quench dynamics.
We consider the quantum Hamiltonian 
\be
 {\cal H} = H(\{\sigma_i^z\})-\Gamma\sum_{i=1}^N\sigma_i^x,
\ee
where $H(\{\sigma_i^z\})$ is the Ising-spin Hamiltonian
as we described in the previous section.
We first prepare state with the ground state of the Hamiltonian 
at $\Gamma\to\infty$.
Then, we set $\Gamma=0$ at $t=0$ and consider the time evolution.
The initial state is given by 
\be
 |\psi\rangle = \prod_{i=1}^N |+x\rangle_i,
\ee
where $|+x\rangle_i$ is the eigenstate of $\sigma_i^x$ with 
the eigenvalue $+1$.
The state evolution at $t>0$ is given by ${\mathrm e}^{-iHt}|\psi\rangle$.
We focus on the return amplitude
\be
 G(t)=\langle\psi|{\mathrm e}^{-iHt}|\psi\rangle.
\ee
$|G(t)|^2$ represents the probability for the state 
to go back to the initial state.
In the $\sigma^z$-basis, the initial state is written as 
\be
 |\psi\rangle 
 = \prod_{i=1}^N \frac{1}{\sqrt{2}}\left(|+z\rangle_i+|-z\rangle_i\right)
 = \frac{1}{2^{N/2}}\sum_{n=1}^{2^N}|n\rangle, 
\ee
where $|n\rangle$ represents one of eigenstates in the $\sigma^z$-basis.
Using this representation, we can write the return amplitude as
\be
 G(t)=\frac{1}{2^N}\sum_{m,n}\langle m|{\mathrm e}^{-iHt}|n\rangle
 =\frac{1}{2^N}\sum_n\langle n|{\mathrm e}^{-iHt}|n\rangle
 = \frac{1}{2^N}Z(\beta=it).
\ee
Thus, the return amplitude is written by using the partition function.
In random systems, we calculate $[|G(t)|^2]=2^{-2N}[Z(it)Z(-it)]$,
which is obtained by setting $n=1$ in the formulation of the previous section.
In the replica calculation, 
the state of the system is characterized by the order parameter
\be
 Q = \left(\begin{array}{cc} 1 & \tilde{q} \\ \tilde{q} & 1 
 \end{array}\right).
\ee
$\tilde{q}$ represents the correlation between 
the system at $\beta=it$ and one at $\beta=-it$.
We see that this order parameter plays the same role as $\tilde{Q}$ 
in the previous calculations of zeros.

\subsection{Mean-field systems}

We examine the random energy model to show that the singularity
appears in the return probability.
In this case, the calculation is easily done by the energy 
representation~\cite{Derrida1, Derrida2}.
As the name of the model suggests, 
the function is written with respect to 
the sum of random energy levels
$G(t)=2^{-N}\sum_{n=1}^{2^N}\exp(-iE_nt)$.
By using the Gaussian random average $[E_n^2]=NJ^2/2$, we find
\be
 [|G(t)|^2] = \frac{1}{2^{N}}
 +\frac{1}{2^{2N}}\sum_{m\ne n} \left[{\mathrm e}^{-i(E_m-E_n)t}\right] 
 \sim \exp\left(-N\ln 2\right)
 +\exp\left(-NJ^2t^2/2\right).
\ee
At the thermodynamic limit, the rate function $g(t)$ defined as 
$[|G(t)|^2]=\exp(-Ng(t))$ is given by 
\be
 g(t)=\left\{ \begin{array}{ll}
 \frac{J^2t^2}{2} & \mbox{for}\ t\le t_{\rm c}=\frac{\sqrt{2\ln 2}}{J} \\
 \ln 2 & \mbox{for}\ t> t_{\rm c}
 \end{array}
 \right..
\ee
The probability freezes at $t=t_{\rm c}$, which 
shows a dynamical singularity.
This transition point $t=t_{\rm c}$ coincides with 
that of the P-DC phase transition in 
the distribution of zeros.
This is not an accidental coincidence since 
the analytical expression is equivalent 
in the spin representation.

We can also solve the case of the Sherrington-Kirkpatrick model.
By using equation (\ref{phiq}) with $p=2$ and $n=1$, we obtain  
\be
 g(t)=\frac{J^2t^2}{2}(1+\tilde{q}^2)-\ln\cosh(J^2t^2\tilde{q}),
\ee
where $\tilde{q}$ is determined by the saddle-point equation 
\be
 \tilde{q} =\tanh(J^2t^2\tilde{q}).
\ee
This has a solution $\tilde{q}>0$ at $t>t_{\rm c}=1/J$.
Thus, we again have a phase transition and the singularity appears in
the rate function.
This result implies that the SK model also has a DC phase.

\begin{figure}[t]
\begin{center}
\includegraphics[width=18pc]{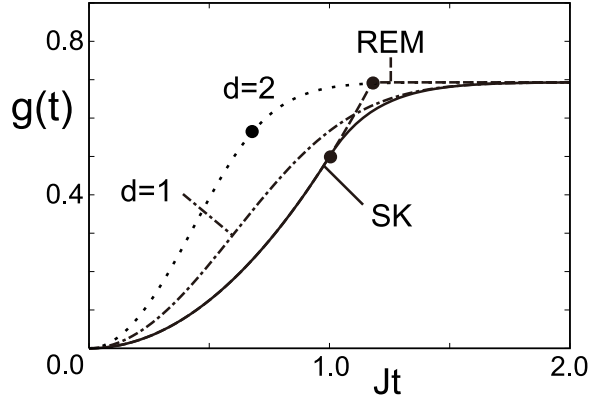}
\caption{\label{fig:gt}The rate function $g(t)$ 
in the random energy model (REM), Sherrington-Kirkpatrick model (SK) and 
Edwards-Anderson model at $d=1$ and $d=2$.
Dots denote singular points.}
\end{center}
\end{figure}

\subsection{Finite-dimensional systems}

We show that the dynamical singularity appears not only in 
mean field models but also in finite dimensional systems.
We use the Edwards-Anderson model defined in equation (\ref{EA}).
After the averaging we obtain the expression
\be
 [|G(t)|^2] = \frac{1}{2^{2N}}{\mathrm e}^{-N_{\rm B}J^2t^2}
 \Tr\exp\left(J^2t^2\sum_{\langle ij\rangle}S_iS^*_iS_jS^*_j\right)
 = \frac{1}{2^{N}}{\mathrm e}^{-N_{\rm B}J^2t^2}Z_{\rm Ising}(J^2t^2),
\ee
where $N_{\rm B}$ is the number of the sum $\langle ij\rangle$
and $Z_{\rm Ising}(\beta)$ is the partition function of the classical 
pure Ising model.
From the result of the Ising model, we can immediately 
conclude that the dynamical singularity occurs at $d\ge 2$ where
$d$ is the dimension of the system.
For example, we can write explicit results 
for the model in one and two dimensional square lattice as
\be
 g(t) = \left\{\begin{array}{ll}
 J^2t^2-\ln\cosh(J^2t^2) & d=1 \\
 2J^2t^2-\ln\cosh(2J^2t^2)
 -\frac{1}{2}\int_0^{2\pi}\frac{{\mathrm d}^2\bm{k}}{(2\pi)^2}
 \ln\left[1-\frac{1}{2}\frac{\sinh(2J^2t^2)}{\cosh^2(2J^2t^2)}
 \left(\cos k_1+\cos k_2\right)
 \right] & d=2 
 \end{array}\right..
\ee
The phase transition at $d=2$ occurs at the critical point defined by 
$2J^2t_{\rm c}^2=\ln(1+\sqrt{2})$.
All of our results on the rate function are 
displayed in figure~\ref{fig:gt}.

For the case of $d=1$, we do not have any singularity in the rate function.
However, this does not mean that there are no zeros 
on the imaginary axis $\beta=it$.
The partition function of the one-dimensional Ising model is written as 
\be
 Z_{\rm Ising}(\beta)=\prod_{i=1}^N
 ({\mathrm e}^{\beta J_i}+{\mathrm e}^{-\beta J_i}).
\ee
This expression shows that the zeros are located 
on the imaginary axis as $\beta J_i=i\pi/2$.
Applying the formula of the density of zeros, we have 
\be
 \rho(\beta_1,\beta_2) = 
 J^2\int\frac{{\mathrm d}z}{\sqrt{2\pi}}{\mathrm e}^{-z^2/2}
 z^2\delta(\sinh\beta_1Jz)\delta(\cos\beta_2Jz),
\ee
for the density of zeros.
Thus, the zeros distribute continuously on the imaginary axis.
However, as we know from the free energy in the spin-glass phase,
the singularity does not necessarily appear on the imaginary axis
since we are on singularities all the way of the evolution.
We expect that the zeros distribute on the imaginary axis 
also in higher dimensional systems.
Therefore, the singularity at $t=t_{\rm c}$
implies that we enter the DC phase 
characterized by two-dimensional distributions of zeros.

\section{Summary}

We developed the theory of zeros in spin-glass systems.
We gave evidence that the chaotic effect is 
relevant to find zeros in the spin-glass phase.
The concept of the replica symmetry breaking is not necessarily
connected to this effect directly.
This is a desirable result 
since the chaotic effect is expected to occur in non-mean-field systems
where the replica-symmetry-breaking ansatz is suspicious to apply.

As an unexpected result, we find zeros in a non-spin-glass phase
in almost all models we studied.
This phase appears around the imaginary axis in the complex-$\beta$ plane.
Since it is far from the real axis, 
the state is not relevant to usual spin-glass transition.
Instead of that, we expect that this chaotic phase is able to explore 
by considering dynamics.
As an example, we propose a quenching protocol of a quantum system.
By relating the quantum amplitude to the partition function at $\beta=it$,
we show that the dynamical singularity is described as a
phase transitions to the chaotic phase.
This occurs not only in mean-field systems
but also in finite dimensional systems.
To the best of our knowledge, this is the first result that 
the spin-glass systems in finite dimensions exhibit 
a dynamical singularity 
even in systems without the thermodynamic phase transition.

We note that our dynamical phase transition is not directly related 
to the one discussed in context of glassy dynamics, at this moment.
As explained in Introduction, the zeros seem to be closely connected 
to change of dominating pure states, but we are not sure that 
this description is useful to discuss the DC phase and 
the related dynamical behavior. 
Clarifying this point will unveil some characteristic behavior 
in quantum quench and possibly lead a connection to 
other dynamical phenomena in glass systems.

We calculated the return probability 
to find the dynamical behaviour of the system.
From a statistical mechanical point of view, 
this calculation corresponds to the annealed average.
It is also possible to consider the quenched average
where the average of the rate function is examined.
Although we consider that 
the annealed calculation is appropriate in the present purpose 
to calculate the return probability, 
it is not clear whether the result from the annealed average 
is generally relevant in characterizing dynamical properties of the system.
This is an interesting problem and is left for future studies.

\section*{Acknowledgments}

T. O. acknowledges the support by the Grant-in-Aid for JSPS Fellows. 


\end{document}